\documentclass[12pt]{article}

\usepackage[table]{xcolor}
\usepackage[normalem]{ulem}
\usepackage{longtable}
\usepackage{footmisc}
\usepackage{orcidlink}

\usepackage{scicite}
\usepackage{ref_macros_sci}

\usepackage{comment}
\usepackage[sectionbib]{bibunits}
\defaultbibliographystyle{sn-nature} 
\defaultbibliography{ms}

\usepackage{cancel}
\usepackage[normalem]{ulem}
\usepackage{graphicx}
\usepackage{setspace}

\usepackage{amsmath}
\usepackage{amssymb}
\usepackage{xspace}
\usepackage{multirow}
\usepackage{array}
\newcolumntype{C}[1]{>{\centering\arraybackslash}p{#1}}
\usepackage{rotating}

\usepackage{siunitx}
\usepackage[flushleft]{threeparttable}

\newcommand\nicer{\textit{NICER}\xspace}

\newcommand{\xmm}{{\it XMM-Newton}\xspace}
\newcommand{\target}{{AT2020ocn}\xspace}
\newcommand{\swift}{\textit{Swift}\xspace}

\newcommand{\ergs}{{\rm erg~s$^{-1}$}\xspace}
\newcommand{\fluxunits}{{\rm erg~s$^{-1}$~cm$^{-2}$}\xspace}

\renewcommand{\figurename}{Figure}

\usepackage[final]{pdfpages}
\usepackage[left]{lineno}

\usepackage{times}
\usepackage[hypcap=false]{caption} 
\usepackage{subcaption, booktabs}

\usepackage{floatrow}
\DeclareFloatVCode{somespace}{\vspace{1.667\baselineskip}}
\usepackage{hyperref}

\newenvironment{sciabstract}{%
\begin{quote} \bf}
{\end{quote}}

\newcounter{lastnote}
\newenvironment{scilastnote}{%
\setcounter{lastnote}{\value{enumiv}}%
\addtocounter{lastnote}{+1}%
\begin{list}%
{\arabic{lastnote}.}
{\setlength{\leftmargin}{.22in}}
{\setlength{\labelsep}{.5em}}}
{\end{list}}

\title{Lense-Thirring Precession after a Supermassive Black Hole Disrupts a Star}
\author{
Dheeraj R. Pasham$^{1}$, Michal Zaja{\v{c}}ek$^{2}$, C.J. Nixon$^{3}$,\\ Eric R.~Coughlin$^{4}$ \orcidlink{0000-0003-3765-6401}, Marzena Śniegowska$^{5}$, Agnieszka Janiuk$^{6}$,\\ Bożena Czerny$^{6}$, Thomas Wevers$^{7,8}$, Muryel Guolo$^{9}$ \orcidlink{0000-0002-5063-0751}, \\Yukta Ajay$^{9}$ \orcidlink{0009-0007-8764-9062}, Michael Loewenstein$^{10}$.
\\
{\small $^{\bf 1}$Kavli Institute for Astrophysics and Space Research,}\\ {\small Massachusetts Institute of Technology, Cambridge, MA, USA}\\
{\small $^{\bf 2}$Department of Theoretical Physics and Astrophysics, Masaryk University, Brno, Czech Republic}\\
{\small $^{\bf 3}$School of Physics and Astronomy, University of Leeds, Leeds, LS2 9JT, UK}\\
{\small $^{\bf 4}$Department of Physics, Syracuse University, Syracuse, New York, USA}\\
{\small $^{\bf 5}$School of Physics and Astronomy, Tel Aviv University, Tel Aviv 69978, Israel}\\
{\small $^{\bf 6}$Center for Theoretical Physics, Polish Academy of Sciences,}\\{\small Al. Lotników 32/46, 02-668 Warsaw, Poland}\\
{\small $^{7}$Space Telescope Science Institute, 3700 San Martin Drive, Baltimore, MD 21218, USA}\\
{\small $^{8}$European Southern Observatory, Alonso de Córdova 3107, Vitacura, Santiago, Chile}\\
{\small $^{\bf 9}$Department of Physics and Astronomy, Johns Hopkins University,}\\{\small 3400 N. Charles St., Baltimore MD 21218, USA}\\
{\small $^{\bf 10}$NASA Goddard Space Flight Center, Greenbelt, MD, USA}\\
}

\begin{document} 
\baselineskip24pt
\maketitle 
\begin{bibunit}
\begin{sciabstract}
An accretion disk formed around a supermassive black hole (SMBH) after it disrupts a star is expected to be initially misaligned with respect to the black hole’s equatorial plane. This misalignment induces relativistic torques (the Lense-Thirring effect) on the disk, causing the disk to precess at early times, while at late times the disk aligns with the black hole and precession terminates. Here, using high-cadence X-ray monitoring observations of a TDE, we report the discovery of strong, quasi-periodic X-ray flux and temperature modulations from a TDE. These X-ray modulations are separated by 17.0$^{+1.2}_{-2.4}$ days and persist for roughly 130 days during the early phase of the TDE. Lense-Thirring precession of the accretion flow can produce this X-ray variability, but other physical mechanisms, such as the radiation-pressure instability, cannot be ruled out. Assuming typical TDE parameters, i.e., a solar-like star with the resulting disk extending at-most to so-called circularization  radius, and that the disk precesses as a rigid body, we constrain the disrupting black hole's dimensionless spin parameter to be $0.05\lesssim |a|\lesssim 0.5$. 
\end{sciabstract}

\target/ZTF18aakelin is an optical transient from the center of a previously-quiescent galaxy at a redshift of 0.0705 \cite{2020ATel13859....1G} (Extended Data Figure, EDF~\ref{fig:sdss}). Follow-up optical spectra taken 1-2 months post-discovery revealed a blue continuum and a broad He~II line (Fig.~16 of \cite{ztftdesample2}). Based on these properties it was classified as a TDE \cite{2020ATel13859....1G,ztftdesample2}. We measured the host galaxy's stellar velocity dispersion using an optical spectrum taken 12 years prior to the outburst (section~\ref{sec:bhmass}). Assuming the scaling relation between the host stellar velocity dispersion and black hole mass implies a disrupting SMBH mass of 10$^{6.4\pm0.6}\, M_{\odot}$, where the errorbar includes both the measurement and the systematic uncertainties in the scaling relation. 

\begin{figure}[htp!]
\begin{center}
\includegraphics[width=0.95\textwidth, angle=0]{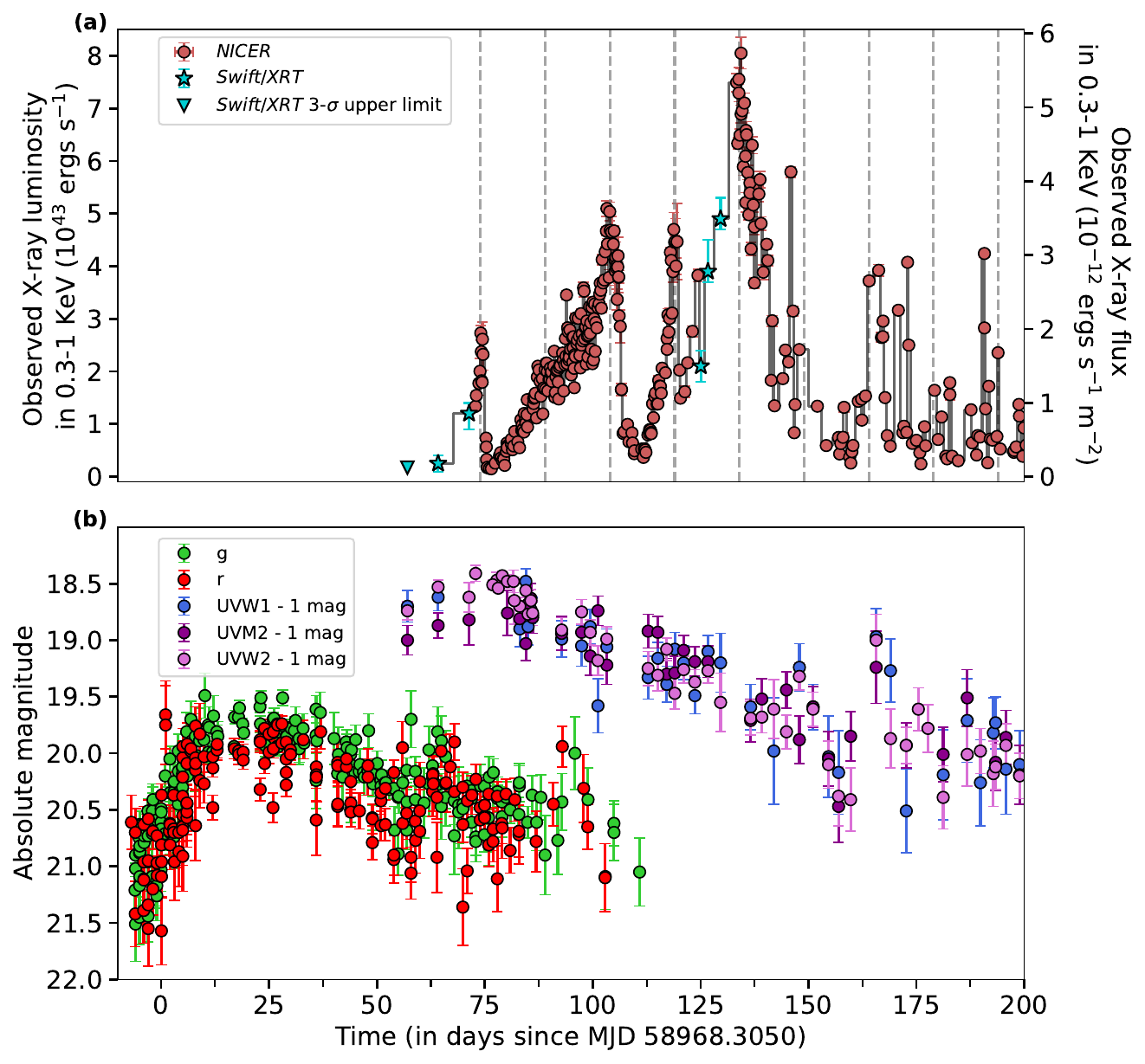}
\end{center}
\caption{{\bf \target's multiwavelength evolution.} {\bf (a) 0.3-1.0 keV X-ray luminosity vs time since optical discovery.} Gaps in \nicer monitoring are filled by \swift data. The dashed/vertical lines are separated by 15 d to guide the eye. Archival \swift X-ray (0.3-1.0 keV) 3$\sigma$ upper limit from prior to MJD 58274 is 3$\times$10$^{-14}$  \fluxunits (4$\times$10$^{41}$ \ergs). The first X-ray/XRT data point is a non-detection with a 3$\sigma$ upper limit of 1.7$\times$10$^{-13}$ \fluxunits. $\textbf{(b) \target's~ optical and UV evolution.}$ All values are host-subtracted. All the other errorbars represent 1$\sigma$ uncertainties. The data are provided as a supplementary file. }\label{fig:fig1}
\end{figure}

Roughly a day after the TDE classification \nicer started a high-cadence (multiple visits per day) monitoring program on MJD 59041. Here we focus on the first $\approx$4 months of monitoring data during which multiple soft X-ray (0.3-1.0 keV) flares are evident (Fig.~\ref{fig:fig1}a and EDF \ref{fig:xrtimage}). A visual inspection suggests that they are regularly spaced, roughly 15-d apart, and similar modulations are not present in the optical/UV bands (Fig.~\ref{fig:fig1}b).

\begin{figure}[htbp!]
    \centering
    \includegraphics[width = \columnwidth]{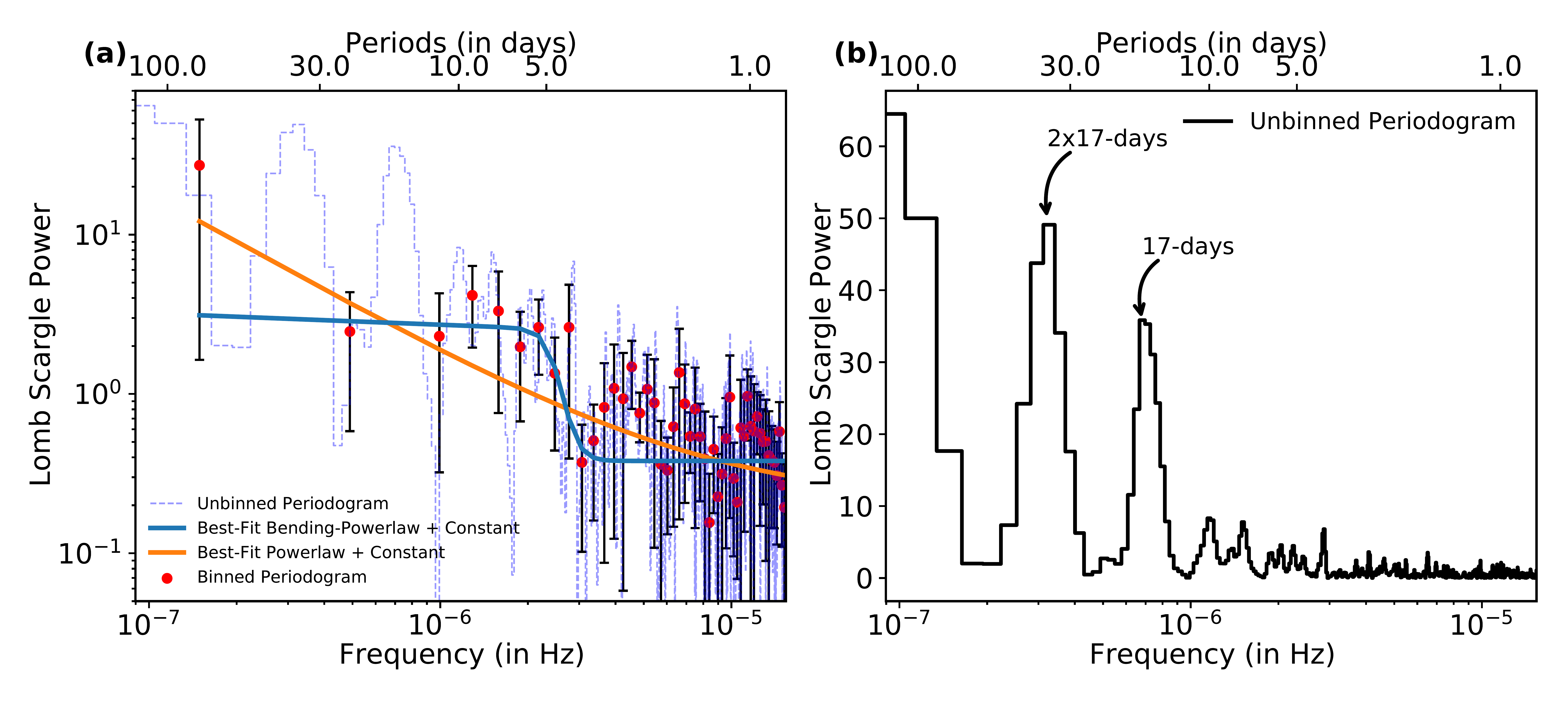}
    \caption{ {\bf Lomb Scargle Periodogram of the observed 0.3-1.0 keV \nicer light curve}. \textbf{(a)} The blue, dashed histogram represents the unbinned LSP while the red data points are the binned LSP excluding the peaks near 15 and 2$\times$15 days. The solid blue and orange curves are the best-fit bending powerlaw + constant and powerlaw + constant models, respectively, used to characterize the noise continuum. \textbf{(b) Unbinned periodogram (same as (a)) but the y-axis is shown on linear scale without overplotting the continuum for clarity. }\label{fig:lsp}}
\end{figure}

To quantify the variability we extracted a Lomb Scargle Periodogram (LSP; \cite{scargle,lspnorm}) of the background-subtracted 0.3-1.0 keV count rate (Fig.~\ref{fig:lsp}). As expected, there is a broad peak at 17$^{+1.2}_{-2.4}$ d with additional harmonics at integer ratios of 17-d. The uncertainty represents the full width at half maximum of the highest bin near 17 days. Given this measurement uncertainty and  the fact that the individual peaks in Fig.~\ref{fig:fig1}a appear better aligned with 15-d vertical lines, we refer to this signal as the 15-d quasi-periodicity throughout the rest of the manuscript. Using a rigorous set of Monte Carlo simulations, we estimate the global statistical significance (false alarm probability/FAP) of finding a broad peak as strong as the one found in the data by chance to be less than 1 in 10,000 (or $>$3.9$\sigma$ for all the continuum models considered, assuming a Gaussian distribution; see details in section~\ref{sec:statsig} and EDFs \ref{fig:whitetests}, \ref{fig:lspbias}, \ref{fig:fap}). Our estimate of the global FAP was 1) tested against a range of underlying noise continuum models, 2) included a search over all frequencies/periods sampled by the LSP (1-100 days), and 3) it included a search for broad peaks in the noise LSPs with a wide range of coherence values ranging from 2-10. Coherence is defined as the ratio of the centroid frequency of a broad power spectral peak over its width and X-ray quasi-periodic oscillations (QPOs) seen in accreting stellar-mass black holes typically have values below 10 \cite{mcclin}.

\begin{figure}[htp!]
\begin{center}
\hspace{-1.35cm}
\includegraphics[width=0.75\textwidth, angle=0]{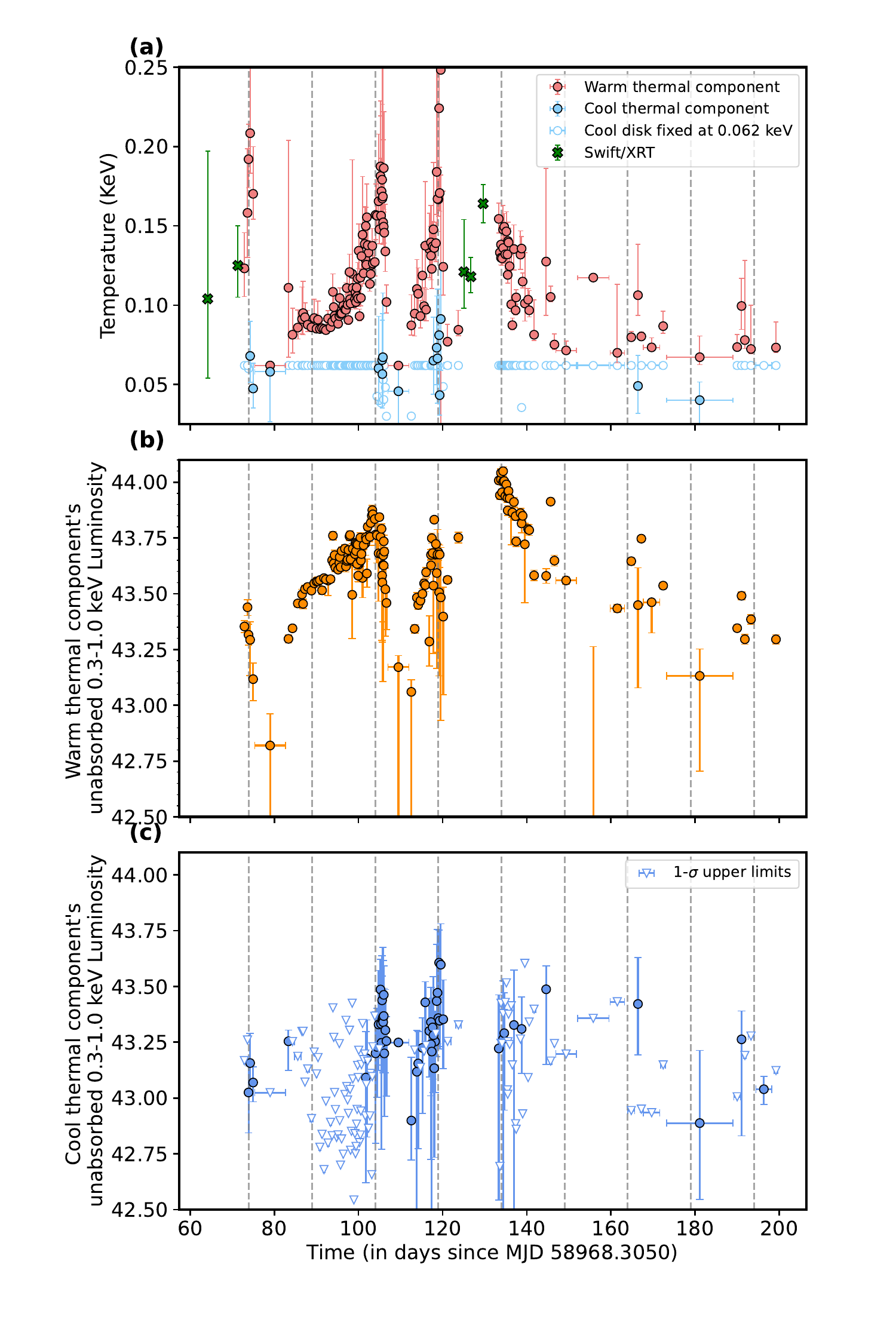}
\end{center}
\vspace{-.35cm} 
\caption{{\bf \target's X-ray spectral evolution.} {\bf (a) Temperature evolution of the warm and the cool 
 X-ray thermal components.} When the cool component was not statistically required by the data, we froze its temperature to 0.062 keV and compute a 1$\sigma$ upper limit on the cool component's flux. Logarithm of luminosity of the warm {\bf (b)} and the cool components {\bf (c)}.  All errorbars represent 90\% uncertainties. }
\label{fig:fig3}
\end{figure}

We then performed time-resolved X-ray spectral analysis on the first 130-days of \nicer data. Our main findings are: 1) \target's X-ray spectrum is soft and can be described by two thermal components: a cool and a warm component, 2) the overall X-ray flux shows quasi-periodic modulations that repeat roughly every 15 days, and 3) these are accompanied by X-ray temperature modulations on the same timescale (Fig.~\ref{fig:fig3}b).

We consider a range of models including a repeating partial TDE \cite{cufari2022}, repeated debris stream self-interactions \cite{clement:streamcollisions,andalman22}, and neutral column density changes, and disfavor them based on \target's observed properties (see supplementary material/SM). Radiation pressure instability (RPI; \cite{1974ApJ...187L...1L,Janiuk:RPIGRS1915}) can reproduce the $\sim$15-d timescale if the outer disk is truncated at roughly 30 gravitational radii, $R_{g}$. However, it predicts a modulation amplitude that is more than two orders of magnitude larger than observed in Fig.~\ref{fig:fig1}a. These amplitudes can be damped to levels comparable to observations if a magnetic field $\sim$10$^{4}$ Gauss is present in the inner disk. Thus RPI can, in principle, explain the X-ray modulations, but requires fine-tuning of multiple parameters (see SM). 

In a TDE the stellar orbit and SMBH spin axis will be misaligned, and disk material will undergo Lense-Thirring precession in which the plane of the orbit precesses around the black hole's spin vector. Emission from the central accretion disk, combined with Lense-Thirring precession of the disk, may provide a straightforward explanation for the soft X-ray spectrum, flux and temperature modulations, and the lack of similar modulations in the optical/UV bands. Indeed, Lense-Thirring precession is commonly accepted to be the cause of some X-ray QPOs of the order of seconds from accreting stellar mass compact objects \cite{Ingram2010,Motta2014,Nixon2014}. The shape of the outbursts we have detected in AT2020ocn are similar to some of those exhibited by the X-ray binaries GRS 1915+105 and IGR J17091–3624. These systems show quasi-periodically repeating state cycles (e.g., \cite{Belloni1997a,Muno1999,Altamirani2011}), and this behavior has often been interpreted as evidence for the radiation pressure instability \cite{Belloni1997,Janiuk:RPIGRS1915}, but it has also been suggested that it may arise from Lense-Thirring precession of a radially narrow region of the disc close to the black hole horizon \cite{Raj2021b}. We therefore consider Lense-Thirring precession of a newly formed accretion disk around a SMBH. This can manifest in two modes: 1) if the accretion rate is sufficiently high, and thus the disk geometrically thick such that the disk angular semi-thickness $H/R$ is larger than the disk viscosity parameter $\alpha \approx 0.1$ \cite{Martin2019}, then the precession can be efficiently communicated via pressure waves \cite{Papaloizou1983}. This allows a significant portion of the disk to precess as a rigid body \cite{2012PhRvL.108f1302S,Franchini2016:TDEprecession}; or 2) for lower accretion rates, where the disk is thin, the inner disk can tear into discrete annuli that precess individually \cite{Nixon2012}. In the former case the observed X-ray modulations are a result of the changing orientation of the system, while in the latter scenario the X-ray modulations result from a combination of changing orientation and accretion of discrete precessing annuli. We focus on the first mode here because the accretion rate is, in general, expected to be high following a TDE (e.g., Fig. 1 of \cite{2018MNRAS.478.3016W}), and discuss the second mode in SM.

When the disk precesses as a rigid body, both the observed flux and the disk temperature can exhibit modulation over the precession period (e.g., see Fig.~4 of \cite{Ulmer1999}). This is because, during certain phases of the precession, our line of sight enables us to observe the hot/inner disk, whereas during other phases, our view of the inner disk becomes obstructed, allowing mostly the outer/cooler disk to be visible (EDF~\ref{fig:scheme}). Also, Fig. \ref{fig:fig3}c suggests that the cool component's variability roughly traces the warm component--albeit with large error bars. This is consistent with precession of an extended disk rather than a narrow ring. While the origin of optical/UV emission from TDEs is still debated, it is, in general, not thought to be direct disk emission. Some current models include stream-stream collisions (e.g., \cite{clement:streamcollisions}) and X-ray reprocessing (e.g., \cite{2013ApJ...767...25G, Roth:TDExrayreprocess}). Thus, a precessing inner disk that produces soft X-rays should not modulate the optical/UV flux originating far away from the hole. Furthermore, theoretical studies have estimated that, for SMBH masses weighing 10$^{5-7}$ M$_{\odot}$, such rigid body precession should last between 0.4--0.7 years before the accretion rate declines to the point that rigid precession is no longer possible and instead the disk aligns with the black hole spin (Fig. 2 of \cite{2012PhRvL.108f1302S}). This is consistent with the observed lifetime of \target's X-ray modulations, $t_{\rm lifetime}$ $\sim$ 130 days (Fig. \ref{fig:fig1}a). \target's first few X-ray flares are asymmetric, and this can be interpreted as a precessing disk that is warped, rather than planar, which can lead to an abrupt obscuration of the inner hot gas. Also, the second peak in the total X-ray flux (close to 90 days in Fig.~\ref{fig:fig1}a) appears blended with the third peak. However, it is more pronounced in the luminosity and the temperature evolution of the warm component (Fig.~\ref{fig:fig3}a,b). You can explain this in the precession model if the disc initially has a significant geometric thickness. This can shield the inner warm material more when compared to other time periods. 


If rigid body disk precession is indeed driving these modulations, we can use the observed period to constrain the disrupting SMBH's spin \cite{2012PhRvL.108f1302S}. We calculated the precession period following \cite{2012PhRvL.108f1302S}, taking the power-law of the surface density profile to be $s = -3/2$, which corresponds to the radiation pressure dominated inner region of the standard disk model \cite{Shakura1973,Novikov1973}. In this case the timescale is principally determined in the outer disk regions and we find that $t_{\rm p} = (\pi/a)\,\left(R_{\rm out}/R_{\rm g}\right)^3\,(GM/c^3)\,\times \xi(R_{\rm in}/R_{\rm out})$, where $a$ is the dimensionless black hole spin, $R_{\rm g} = GM/c^2$ is the gravitational radius of the black hole, and $\xi(R_{\rm in}/R_{\rm out})$ is a dimensionless function that is weakly dependent on the ratio of the inner and outer disk radii (for $R_{\rm in}/R_{\rm out} \rightarrow 0$, we have $\xi \rightarrow 1/4$).
Using this approximation, taking $R_{\rm out}$ to be the circularisation radius of the debris, and inverting the equation for the precession timescale, we can calculate the black hole spin as
\begin{equation}
a \approx 0.1 \left(\frac{t_{\rm p}}{15.9\,{\rm days}}\right)^{-1}\,(\beta/2)^{-3}\,(\xi/0.5)\,M_{\star,\odot}^{-1}\,R_{\star,\odot}^{3}M_7^{-1}\,,
\end{equation}
where $\beta = R_{\rm t}/R_{\rm p} \simeq 2$ is the impact parameter of the stellar orbit required to just fully disrupt a solar-like star \cite{Guillochon2013}, $M_{\star,\odot}$ and $R_{\star,\odot}$ are the mass and radius of the star in solar units, and $M_7$ is the SMBH mass scaled by $10^7\,M_\odot$.

\begin{figure}[htp!]
\begin{center}
\includegraphics[width=\textwidth, angle=0]{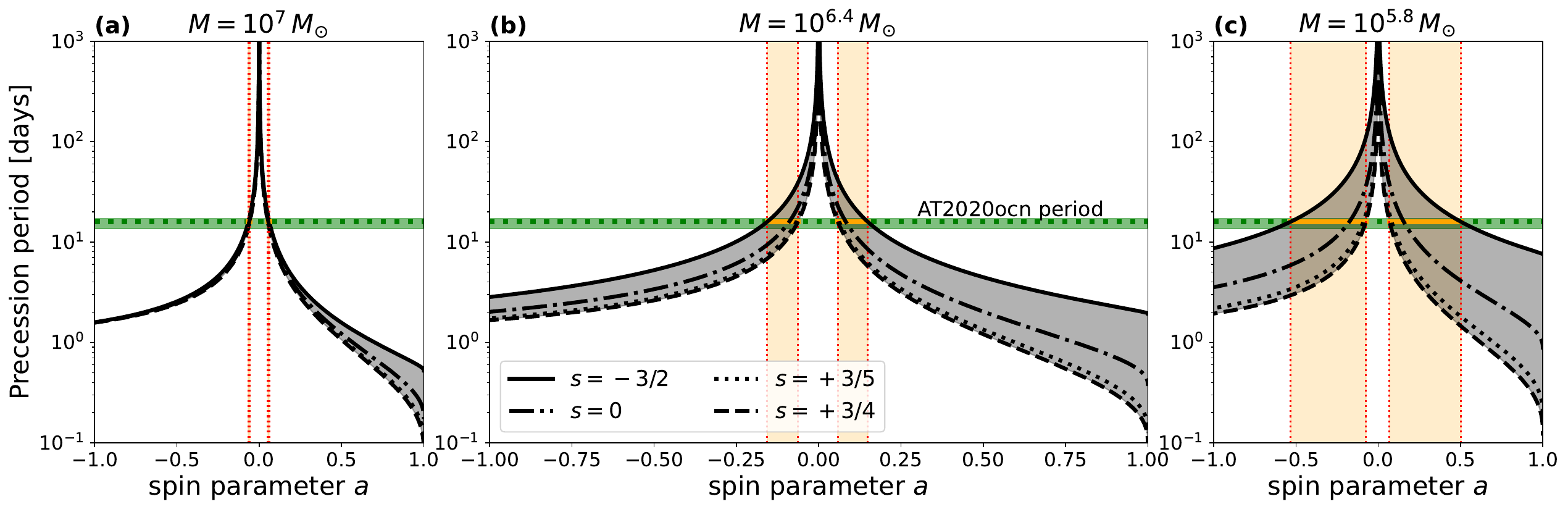}
\end{center}
\caption{{\bf Spin constraints on \target's SMBH based on the rigid-body Lense-Thirring precession model}. Panels (a), (b) and (c) show the precession period-spin relations for the upper-limit, best-fit and the lower-limit on the SMBH mass inferred from the M-$\sigma_{*}$ relation (section \ref{sec:bhmass}). Standard TDE parameters, i.e., a solar-like star with the resulting disk extending out to tidal radius, were assumed for these calculations. The shaded grey area captures the uncertainty arising from various disk density profiles. The green dashed horizontal line represents \target's rest-frame period ($15.9$ days) and the green band reflects the uncertainty in the rest-frame period ($15.9^{+1.1}_{-2.2}$ days). For a given SMBH mass, the possible spin values are depicted by the orange-shaded area. The overall spin constraint for \target's SMBH across all masses is $0.05\lesssim |a|\lesssim 0.5$ (see the Extended Data Table \ref{tab:tab1} for specific values). }
\label{fig:fig4}
\end{figure}

From this equation we can see that to accurately constrain the spin with this model we require accurate constraints on the parameters $\beta$, $M_{\star,\odot}$, $R_{\star,\odot}$ and $M$. However, if we take standard TDE parameters (as above), and an SMBH mass estimate of $10^{6.4}\,M_\odot$, then we find that $a \approx 0.15$ is required to generate the 15\,day period. We also explored varying the surface density profile's power-law index, $s$, in the range $-3/2$ to $3/4$, finding that for SMBH masses $\sim 10^7M_\odot$ the value of $s$ makes little difference to this estimate, while at the lower end of the black hole mass range the value of $s$ can make a significant difference with $s = -3/2$ providing the largest spin estimate. Thus considering a range of possible models, we conclude that $0.05 \lesssim \left|a\right| \lesssim 0.5$ (Fig.~\ref{fig:fig4}). Future
modelling of \target's multi-wavelength emission, especially the high-cadence X-ray observations with numerical simulations, will provide a more detailed understanding of the accretion flow structure and should provide even tighter constraints on the SMBH spin. 

The accretion of stellar debris following a TDE can be highly chaotic owing to various physical effects, and this is evident in the plethora of low-cadence X-ray TDE light curves (e.g., \cite{2017ApJ...838..149A,Guolo2023}). Our work demonstrates that high-cadence X-ray monitoring can unveil regular processes happening amongst the chaos from the highly-relativistic regions close to the disrupting SMBHs. If regular modulations like the ones identified here are indeed caused by rigid body disk precession, this enables an independent avenue to measure SMBH spins. Especially exciting is the prospect that upcoming all-sky surveys like the {\it Rubin} observatory could detect hundreds of TDEs per year (e.g., \cite{2020ApJ...890...73B}). Even if only a fraction of them show early X-rays and quasi-periodic precession-like modulations, this could result in independent constraints on SMBH spin distribution in the local Universe.


\clearpage
\vfill\eject


\newpage


\newpage
\vfill\eject

\clearpage
\begin{scilastnote}
\begin{sloppypar}
\item[] {\bf Acknowledgments.}\\ 
MZ acknowledges the GA\v{C}R Junior Star grant GM24-10599M for financial support. CJN acknowledges support from the Science and Technology Facilities Council (grant number ST/Y000544/1) and the Leverhulme Trust (grant number RPG-2021-380). ERC acknowledges support from the National Science Foundation through grant AST-2006684 and the Oakridge Associated Universities through a Ralph E.~Powe Junior Faculty Enhancement award.

{\bf Author contributions:} D.R.P led the overall project, acquired X-ray data, performed reduction and wrote a significant portion of the paper. 

{\bf Competing interests:} The authors declare that there are no competing interests. 

{\bf Data Availability:}
All the {\it NICER}, \xmm and {\it Swift} data presented here is public and can be found in the NASA archives at the following URL: \url{https://heasarc.gsfc.nasa.gov/cgi-bin/W3Browse/w3browse.pl}. Data shown in Figures 1 and 2 can be found on Zenodo \url{https://zenodo.org/records/10062826}. \xmm spectra are available at \url{https://doi.org/10.5281/zenodo.8252931}. Time resolved \nicer spectra can be downloaded from \url{https://doi.org/10.5281/zenodo.8253537}. \\

\end{sloppypar}
\end{scilastnote}

\clearpage
\putbib
\end{bibunit}

%
%
\newpage
\setcounter{page}{1}
\renewcommand{\theequation}{S\arabic{equation}}
\section*{{\Huge Methods.}}
\begin{bibunit}

\section{\Large{\bf Observations and Data Analysis}}\label{supsec:data}
We used data from the Neutron Star Interior Composition ExploreR (\nicer; X-ray), \xmm (X-ray), \swift (X-ray and UV), the Zwicky Transient Facility (ZTF; optical; \cite{ztf}) and Sloan Digital Sky Survey (SDSS; optical). In the following sections, we describe these data and their reduction procedures. Throughout, we adopt a standard $\Lambda$CDM cosmology with H$_{0}$ = 67.4 km~s$^{-1}$~Mpc$^{-1}$, $\Omega_{m}$ = 0.315 and $\Omega_{\Lambda}$ = 1 - $\Omega_{m}$ = 0.685 \cite{planck}. Using the Cosmology calculator of \cite{Wright2006}, \target's redshift of 0.0705 corresponds to a luminosity distance of 330~Mpc.

\subsection{X-ray data}

\subsubsection{\nicer}\label{supsec:nicer}
{\it NICER} started monitoring \target on 2020 July 11, roughly 10 weeks after ZTF discovered it on 29 April 2020 (MJD 58968.305) \cite{2020ATel13859....1G}. {\it NICER} continues to monitor AT2020ocn on a daily cadence at the time of writing of this paper. However, for this work, we only consider data taken over the first $\approx$ 130 days, i.e., until 2020 November 17, when X-ray flares are prominent. 

\begin{sloppypar}

We started our analysis by downloading the raw, level-1, publicly available data from the HEASARC archive: \url{https://heasarc.gsfc.nasa.gov/cgi-bin/W3Browse/w3browse.pl} and reduced them using the {\tt nicerl2} tool with the default screening filters recommended by the \nicer data analysis guide: \url{https://heasarc.gsfc.nasa.gov/docs/nicer/analysis\_threads/nicerl2/}. The resulting good time intervals (GTIs), i.e., uninterrupted data segments, ranged from 250 to $\approx$2600 seconds.

\nicer is comprised of 52 co-aligned detectors (Focal Plane Modules/FPMs). In any given GTI, some detectors may be ``hot'', i.e., optical light loading can make them behave anomalously. We identify such ``hot'' detectors on a per GTI basis by flagging FPMs whose 0.0-0.2 keV count rate is more than 3$\sigma$ above the sigma-clipped median value of all FPMs during that GTI (see \cite{2023NatAs...7...88P,2018cowpasham} for more details).

Using the 3c50 background model \cite{Remillard2022}, we first  extracted the source and the background spectra on a per GTI basis by excluding the appropriate hot FPMs. We then computed the background-subtracted count rates (counts s$^{-1}$ FPM$^{-1}$). Because the source is ``faint'' as per \cite{Remillard2022}, we also applied the so-called level-3 filtering as described in \cite{Remillard2022}. The source was typically above the estimated background in the 0.3-1.0 keV band. Therefore, we adopted this energy for the majority of our analysis in this paper. However, because both the source and the  background are variable, in some GTIs, the background exceeded the source counts down to 0.6 keV. For those spectra, modeling was restricted to an appropriate value lower than 1 keV. 

All spectra were binned using an optimal binning scheme of \cite{Kaastra2016} to have at least 25 counts per spectral bin. This was done using the {\tt ftgrouppha} task of HEASoft: \url{https://heasarc.gsfc.nasa.gov/lheasoft/help/ftgrouppha.html}. We performed spectral modeling in {\it XSPEC} \cite{Arnaud1996} and {\tt Python} version of {\it XSPEC}, {\it PyXspec}, using the $\chi^{2}$ statistic.

\end{sloppypar}

\subsubsection{\xmm}
\xmm observed \target on four occasions, two taken roughly a week after \nicer started monitoring the source (2020 July 18 and 21) and the other two taken 1-2 years later. Here we only use the data from EPIC-pn detector from the first two datasets with observation IDs: 0863650101 (XMM\#1) and 0863650201 (XMM\#2). We reduced the publicly available raw data using the standard {\it epproc} tool of XMM software {\tt xmmsas version 19.1.0} with the latest calibration files. From these cleaned eventfiles we visually inspected the 10-12 keV count rates from the entire field of view to identify epochs of background flaring. GTIs were chosen to exclude these flaring windows. Using only the events that occurred during the GTIs, we extracted source spectra (corrected for pileup) by using annuli with inner radii of 5$^{\prime\prime}$ and 10$^{\prime\prime}$ for XMM\#1 and XMM\#2, respectively. The annuli were centered on optical coordinates: (13:53:53.803, +53:59:49.57) J2000.0 and had an outer radius of 20$^{\prime\prime}$. Background spectra were extracted using events from two nearby circular regions of 50$^{\prime\prime}$ each. Finally, spectra were grouped using {\it xmmsas} tool {\it specgroup} to have a minimum of 1 count per spectral bin.

\subsubsection{\swift}
\swift started monitoring \target on 2020 June 25 at a much lower cadence than \nicer, roughly one visit every few days. There were also 14 archival observations with a total exposure of 12.3 ks prior to the outburst, i.e., between MJD 57255.452 (2015 Aug 14) and MJD 58268.978 (2018 May 30). Using the standard {\tt xrtpipeline} tool we reduced all the X-Ray Telescope (XRT) observations taken prior to 2020 Nov 17. Source events were extracted from a circular aperture of 47.1$^{\prime\prime}$ and background events were chosen from an annulus with an inner and outer radii of 70$^{\prime\prime}$ and 210$^{\prime\prime}$, respectively. We only used events with grades 0-12 as recommended by the data analysis guide. \swift data was used for three reasons: 1) to estimate an upper limit on the X-ray flux prior to the outburst, 2) to fill-in \nicer data gaps, and 3) to confirm that there are no contaminating sources within \nicer's field of view of \target (EDF \ref{fig:xrtimage}).

\subsection{Optical/UV Observations}
\subsubsection{Zwicky Transient Facility}
\target was discovered and reported by ZTF and released as a transient candidate ZTF18aakelin
in the Transient Name Server \cite{Frederick_TNS}. We performed point spread function (PSF) photometry on all publicly available ZTF data using the ZTF forced-photometry service \cite{Masci_2019} in $g$- and $r$-band. We report our photometry, corrected for Galactic extinction of $A_{\rm V}=0.0153$~mag \cite{Schlafly_2011}.

\subsubsection{\swift/UVOT}\label{sec:UVOT}
We perform photometry on \swift/UVOT \cite{Roming_2005} observations of \target\ with the \textit{uvotsource} task in HEAsoft package v6.29 using a 5$^{\prime\prime}$\ aperture on the source position. Another nearby region of 40$^{\prime\prime}$ free of any point sources was used to estimate the background emission. The host contribution was subtracted using a modeled spectral energy distribution (SED). Similar to ZTF data, UVOT photometry was corrected for Galactic extinction.

\section{Black Hole Mass from Host Galaxy's Stellar Velocity Dispersion}\label{sec:bhmass}
\target's host galaxy was observed by SDSS on 2008 Feb 12, i.e., $\approx$12 years before the flare occurred. No narrow emission lines are visible in the spectrum, indicating that the host is a quiescent galaxy. We divide the flux-calibrated spectrum by the median flux value to quasi-normalize the spectrum. Then, we use the penalized pixel fitting routine \cite{Cappellari17} combined with the MILES single stellar population template library \cite{Falcon-Barroso11} to measure the velocity dispersion of the stellar absorption lines ($\sigma_{*}$). We conservatively mask the locations of prominent emission lines during this process (even though none are apparent). Following \cite{Wevers17} we resample the spectrum within the errors and repeat the fitting procedure 1000 times, and take the mean and standard deviation as the velocity dispersion $\sigma_{*}$ and its uncertainty. We find $\sigma_{*}$ = 82$\pm$4 km s$^{-1}$, which translates into a black hole mass of log$_{10}$(M) = 6.4$\pm$0.6 M$_{\odot}$ using the M--$\sigma_{*}$ relation of \cite{Mcconnell13}. Here we have added the measurement uncertainty in quadrature with the systematic uncertainty in the M--$\sigma_{*}$ relation.


\section{Estimating the Statistical Significance of the 15-d X-ray Flux Modulations}\label{sec:statsig}
We employ the following procedure to estimate the \emph{global} statistical significance of the 15-d quasi-periodicity seen in the X-ray light curve in Fig. \ref{fig:fig1}a. The main steps are:
\begin{enumerate}
    \item Estimate the nature of the continuum noise in the periodogram.
    \item Simulate a large number of light curves that follow the above continuum variability.
    \item Sample these simulated light curves exactly as the real/observed data.
    \item Compute the Lomb Scargle Periodograms (LSPs) exactly as done on real data and from these estimate the likelihood of finding a quasi-periodic oscillation (QPO) as strong as the one found in real data with a wide range of coherence values. 
\end{enumerate}

We describe each of these steps in detail below.

\subsection{Estimating the nature of the continuum in the periodogram}
A quick visual inspection of the X-ray light curve (Fig. \ref{fig:fig1}a) suggests that there are 7-8 prominent flares roughly 15 days apart (see vertical dashed lines in Fig. \ref{fig:fig1}a to guide the eye). The second peak appears to be blended with the third one. These regular flares terminate beyond MJD 59171 as a corona is eventually formed and those observations are discussed in a separate paper. We started our timing analysis by computing the LSP of the observed, background-subtracted 0.3-1.0 keV X-ray light curve using all the data until MJD 59171. The LSP was computed exactly as described in \cite{scargle} and normalized as per \cite{lspnorm}. As expected, there is a broad peak in the LSP around 15 days. There are also broad peaks in the LSP at integer harmonics, i.e., 2$\times$15-d, 4$\times$15-d and smaller peaks near $\frac{1}{2}\times$15-d and $\frac{1}{4}\times$15-d (Fig.~\ref{fig:lsp}). To assess the nature of the noise, i.e., the distribution of the power values within the LSP, we used the Kolmogorov-Smirnov and Anderson-Darling tests for white noise. To assess the nature of the LSP in the vicinity of $\sim$15-d, we used the power values that correspond to timescales longer than 3 days and excluded bins near 15-d, 2$\times$15-d, and 4$\times$15-d. Following the procedure outlined in section 2.2.2 of \cite{2018cowpasham}, both the Kolmogorov-Smirnov and Anderson-Darling tests suggest that the null hypothesis that these LSP powers (normalized by mean) below 3 days are white cannot be rejected at even the 90\% confidence level (EDF~\ref{fig:whitetests}). 

The average of the power values corresponding to timescales longer than 3 days (excluding those near 15, 2$\times$15 and 4$\times$15-d) is elevated compared to the average value for shorter than 3 days. This could be due to two reasons: 1) the contribution from the wings of the broad peaks at 15-d and its harmonics, or 2) genuine red noise. Therefore, we compute the FAPs separately for both these cases. 

Red noise is a common type of variability noise that is predicted from both general relativistic magnetohydrodynamic simulations (see, for example, \cite{2020MNRAS.497.1066M}) and X-ray observations and is, in general, described analytically by a power-law or a bending powerlaw \cite{2003ApJ...593...96M}. We rebinned the LSP by a factor of 10 and fit it with a powerlaw + constant and a bending powerlaw + constant models. The former and the latter yielded a $\chi^{2}$/dof of 26/48 and 24/46, respectively. For the powerlaw model, the best-fit index,  normalization and constant values were 1.0$\pm$0.3,  (1.2$^{+4.4}_{-1.2}$)$\times$10$^{-6}$, and 0.2$\pm$0.1,  respectively. For the bending powerlaw model the best-fit normalization, low-frequency powerlaw index, bend frequency, high-frequency index and constant values were 0.7$\pm$5.0, 0.1$\pm$0.5, (2.5$\pm$0.3)$\times$10$^{-6}$ Hz, 15$\pm$13, and 0.38$\pm$0.03, respectively. Below, we describe our analysis for the powerlaw + constant models but we repeat the same procedure for the bending powerlaw case. 

It is known that the best-fit powerlaw index value inferred from modeling the LSP can be  biased if the time series is unevenly sampled \cite{psresp}. To test whether the current sampling could have biased our estimate of the index we carried out the following tests. 

First, using the algorithm of \cite{timmer}, we simulated 10,000 time series whose power spectrum is defined by the best-fit powerlaw + constant model of the real data, i.e., an index of unity and a normalization of 1.2$\times$10$^{-6}$. These time series had a resolution of 100 s. To account for red-noise leakage \cite{psresp} we ensured each of these time series were 10 times longer than the observed baseline of $\sim$130 days. Then, we sampled each of these 10,000 time series exactly as the observed light curve and computed their LSPs. Then we rebinned them by a factor of 10--similar to the observed LSP--and fit them with a powerlaw + constant model. The best-fit powerlaw index values had a median and standard deviation of 0.93 and 0.07, respectively (EDF~\ref{fig:lspbias}), i.e., consistent with the best-fit values shown in Fig.~\ref{fig:lsp}. This demonstrates that, for the current uneven sampling, the inferred power-law index from modeling the LSP does indeed represent the true shape of the underlying power spectrum. 

\subsection{Monte Carlo Simulations of Time Series}
Based on the analysis in the above section we concluded that the underlying continuum can be described as white, a powerlaw or a bending-powerlaw/red noise. The goal in this section is to answer the following question: how often would we see a broad QPO-like feature as strong as the one seen in the observed LSP for each of these underlying noise continuum models. To address this question we use the following methodology for six continuum models: 1) white noise, 2) best-fit powerlaw red noise model corresponding to (index, normalization) = (1.0, 1.2$\times$10$^{-6}$), 3) red noise model corresponding to (index, normalization) = (best-fit index + 1$\sigma$ error, corresponding normalization) = (1.3, 1.0$\times$10$^{-7}$), 4) a red noise model with (corresponding index, best-fit normalization + 1$\sigma$ uncertainty) = (0.9, 5.6$\times$10$^{-6}$), 5) a bending-powerlaw red noise model corresponding to the best-fit parameters, and 6) a bending-powerlaw noise corresponding to the best-fit normalization + 1$\sigma$ uncertainty. 

\begin{enumerate}
\item Using the algorithm of \cite{timmer}, we simulated 10,000 LSPs sampled exactly as the real data and compute its median, LSP$_{median}$.
    \item We then normalize each of the 10,000 LSPs with LSP$_{median}$.
    \item Find all QPO-like features with coherence, Q, between 2 and 10. This was automated by carrying out a sliding window cross-correlation with a Lorentzian whose width at a given frequency is defined by Q. 
    \item Compute the sum of powers, $\Sigma_{p}$, over the width of these features and save the maximum of these sums, $\Sigma_{p,max}$, i.e., the strongest QPO-like feature.
    \item Plot a cumulative distribution function (CDF) of $\Sigma_{p,max}$ values (EDF~\ref{fig:fap}).
    \item Run steps 1-3 from above on the observed LSP to compute the $\Sigma_{p,max, observed}$ and overlay its position on the CDF from step 4. In all the cases, this step found the peak near 15-days. 
\end{enumerate}

Based on this analysis we concluded that the QPO feature near 15-d is statistically acceptable (EDFs~\ref{fig:fap}).

\begin{sloppypar}
\section{X-ray Spectra Analysis}
\subsection{Preliminary X-ray Spectral Modeling with \xmm/EPIC-pn data}
We started our spectral modeling with XMM\#2 which had 3214 counts in the 0.3-1.2 keV band. Because the spectrum was soft we first fit it with a single thermal component, {\it tbabs*ztbabs*zashift*{diskbb}} in {\it XSPEC} leaving all but {\it tbabs} parameters to be free to vary. The MilkyWay column density was fixed at 1.3$\times$10$^{20}$ cm$^{-2}$ using \url{https://heasarc.gsfc.nasa.gov/cgi-bin/Tools/w3nh/w3nh.pl}. This gave a poor fit with C-statistic/degrees of freedom (dof) of 44.0/19. Strong systematic residuals above 0.6 keV were evident.  Next, we added a powerlaw component which yielded a better C-stat/dof of 25.8/17. However, the best-fit powerlaw index was extreme with a value of 5.7$^{+1.2}_{-2.1}$. For comparison, typical index values for persistently accreting SMBHs, i.e., AGN, are below 2 with a value near 3 considered to be extreme \cite{dadina08}. Index values $\sim$6 are unphysical because they imply an unrealistically high intrinsic luminosity when extrapolated to lower energies. Such steep index values can be explained by the fact that in the narrow bandpass of 0.3-1.2 keV we are fitting the Wien's portion of a thermal component, which naturally leads to a steep index when modeled with a power-law. Therefore, we next tried fitting two thermal components, i.e., {\it tbabs*ztbabs*zashift(diskbb+diskbb)} in {\it XSPEC}. This resulted in C-statistic/dof of 22.9/17. The two temperatures were 0.064$^{+0.008}_{-0.010}$ keV and 0.119$^{+0.068}_{-0.027}$ keV (see EDF.~\ref{fig:fig2}).  We also tried {\it tbabs*ztbabs*zashift(diskbb+blackbody)} model which resulted in a similar C-statistic/dof value of 22.8/17. In all these cases, the fit required the neutral column density of the host, i.e., {\it ztbabs}, to be 0. We repeated the same analysis on a few of the \nicer spectra from near the peaks of the flares in Fig.~\ref{fig:fig1}a and concluded that two thermal components describe the data better than a single thermal component. The exact choice of for the thermal components, i.e., {\it diskbb} versus {\it blackbody}, does not matter for the overall conclusions.

\subsection{Time-resolved X-ray spectral modeling using \nicer data}
Motivated by the above spectral modeling and recent detections of two temperature X-ray spectra (e.g., \cite{2019Natur.573..381M,2022ApJ...925...67L}) we adopted a model with two thermal components. To track the two components over the first $\sim$4 months we extracted composite \nicer spectra by combining neighboring GTIs to have at least 1000 counts in each spectra in the 0.3-1.0 keV and more than 50 counts in the 0.75-1.0 keV band. This resulted in 165 spectra between MJDs 59041 and 59171 with median (standard deviation) counts of 3700 (2400). These were fit separately with a single disk blackbody {\it tbabs*zashift*(diskbb)} in {\it XSPEC} and two disk blackbodies {\it tbabs*zashift(diskbb + diskbb)}. In all the cases, the column density near the host ({\it ztbabs}) was pegged by the fit to 0. For each spectrum, we computed the evidence ratio as per the Akaike Information Criterion (AIC) for two disks with respect to a single disk model. If the evidence ratio was less than 10, we fixed the cool disk to a value of 0.062 keV and estimated the cool disk's luminosity. On the other hand, if the evidence ratio was greater than 10, the cool thermal component's temperature was allowed to be free (see Fig.~\ref{fig:fig3}).

\subsection{X-ray flux variations are not driven by neutral column density changes near the TDE}
To test whether the observed changes in X-rays are driven by changes in column density, we extracted higher count \nicer spectra from near the peaks of the early time flares and compared them with spectra between the flares by letting the neutral column density be a free parameter. In all these cases, the best-fit column density of the host was again close to 0. For instance, a composite \nicer spectrum using data taken between MJDs 59073.54 and 59074.45 had roughly 32,200 counts in the 0.3-1.0 keV band. Modeling this spectrum with {\it tbabs*ztbabs*zashift*(diskbb+diskbb)} yielded a best-fit column of the {\it ztbabs} component to be close to 0. For comparison, the 0.3-1.0 keV flux during this time was $\sim$3$\times$10$^{-12}$ \fluxunits which is a factor of 6 and 4 higher compared to XMM\#1 and XMM\#2 epochs. Another example is a spectrum obtained using data between MJDs 59087.085 and 
 59087.877 which had about 21,650 counts. The source flux during this epoch was 3.5$\times$10$^{-12}$ \fluxunits and again the best-fit {\it ztbabs} column was close to 0. Based on these tests, we concluded that the observed changes in X-ray flux were not driven by changing neutral column near the TDE.

\end{sloppypar}

\clearpage
\renewcommand{\figurename}{Extended Data Figure}
\setcounter{figure}{0}

\clearpage
\begin{figure}[htbp!]
    \centering
    \includegraphics[width = \columnwidth]{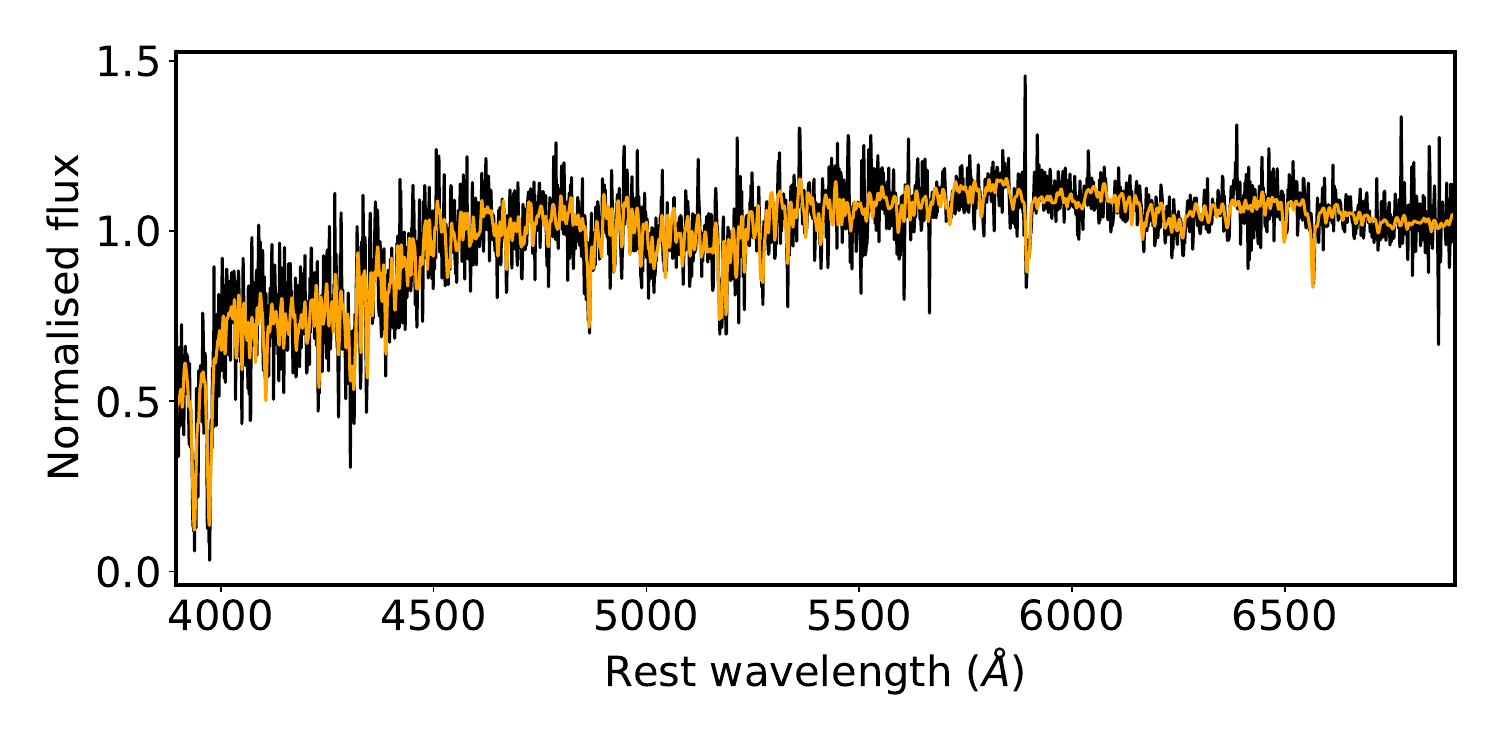}
    \caption{SDSS host spectrum of AT2020ocn (black). Overlaid in orange is the best fit MILES model template obtained through pPXF fitting, broadened to a velocity dispersion of 82$\pm$4 km s$^{-1}$. No narrow emission lines are evident in the spectrum. \label{fig:sdss}}
\end{figure}

\begin{figure}[htbp!]
    \centering
    \includegraphics[width = 0.75\columnwidth]{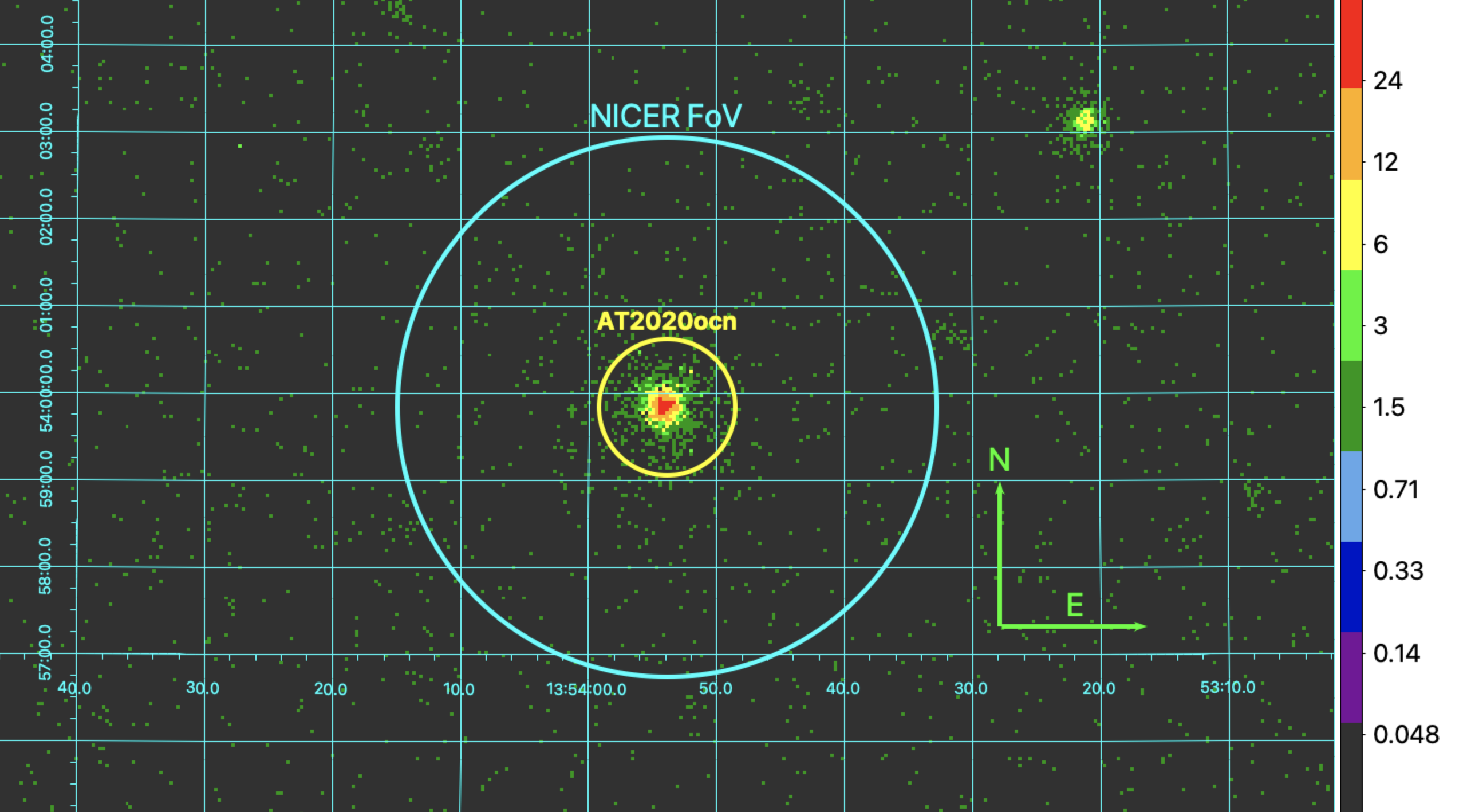}
    \caption{ {\bf \swift/XRT 0.3-1.0 keV image of \nicer's FoV}. The yellow  circle with a radius of 47$^{\prime\prime}$ and is centered on \target's optical coordinates of 13:53:53.803, +53:59:49.57 (J2000.0 epoch). The outer  cyan circle shows {\it NICER}/XTI's approximate field of view of 3.1$^{\prime}$ radius. There are no contaminating sources within {\it NICER}'s field of view.  The north and east arrows are each 100$^{\prime\prime}$ long. The colorbar shows the number of X-ray counts. \label{fig:xrtimage}}
\end{figure}

\begin{figure}[htp!]
\begin{center}
\includegraphics[width=\textwidth, angle=0]{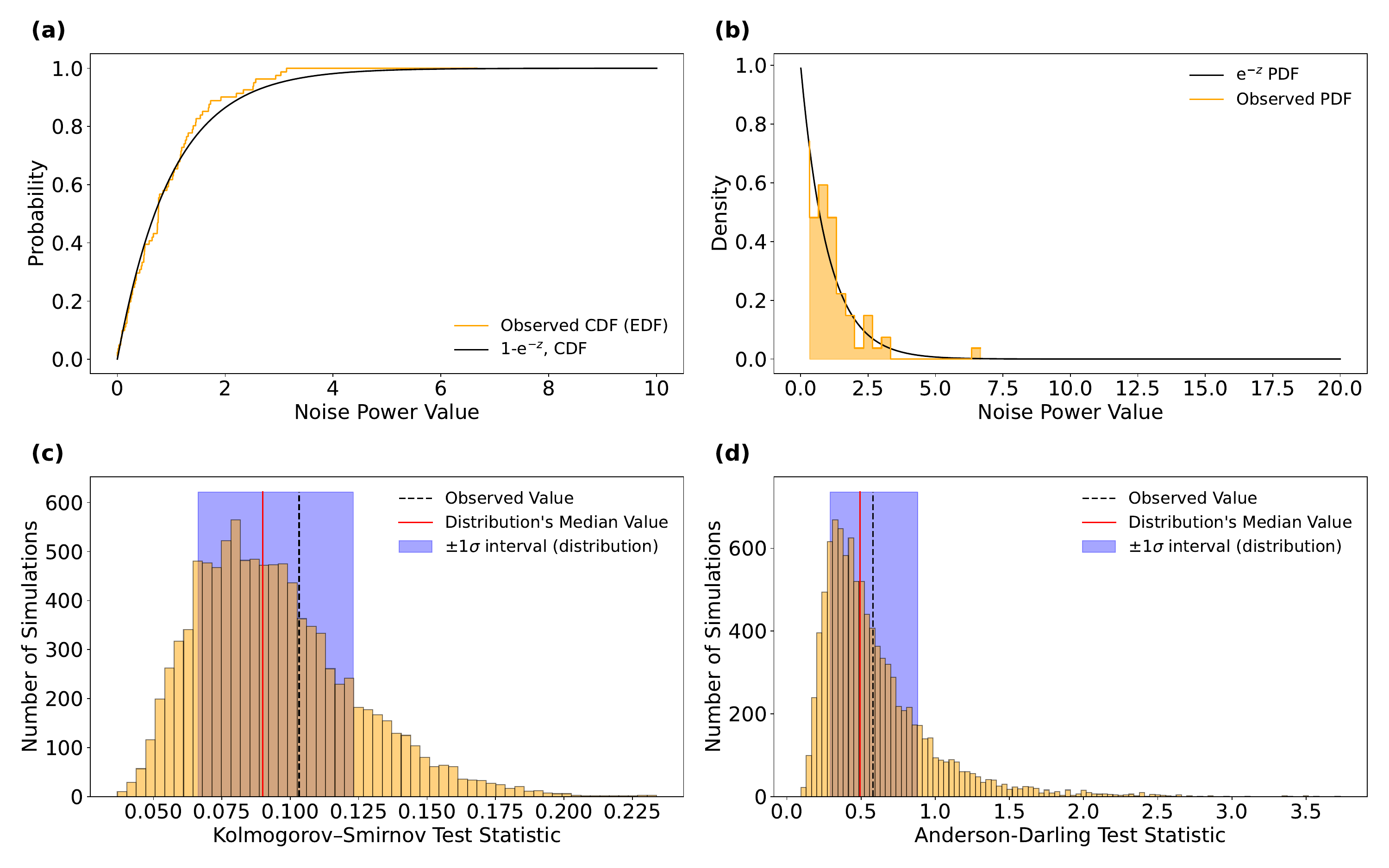}
\end{center}
\vspace{-.35cm} 
\caption{{\textbf{White noise tests for the distribution of noise powers in the observed Lomb Scargle periodogram (LSP) using power values that correspond to timescales slower than 3 days and excluding bins near 15, 2$\times$15 and 4$\times$15 days.}} \textbf{(a) Comparison of the cumulative distribution functions (CDFs) of the observed noise powers and the expected exponential distribution.} The CDF of the observed LSP of the observed light curve is represented by the orange histogram, while the solid black line shows the CDF of white noise powers, which follows an exponential distribution. \textbf{(b) Comparison of the probability density functions of the observed noise powers and the expected exponential distribution.} The solid black line is the PDF of white noise distribution, whereas the shaded orange histogram represents the PDF of the observed noise powers in the LSP of the observed light curve.\textbf{(c) Distribution of the Kolmogorov-Smirnov test statistic derived from simulations}. \textbf{(d) Distribution of the Anderson-Darling test statistic using simulations}. In both {\bf (c)} and {\bf (d)}, the solid red and the dashed black lines represent the median of the distribution and the observed value, respectively and the shaded blue regions indicate the $\pm$1$\sigma$ values of their respective distributions. The null hypothesis that the noise powers are exponentially distributed, i.e., consistent with white noise, cannot be rejected.} 
\label{fig:whitetests}
\end{figure}
%

\begin{figure}[htbp!]
    \centering
    \includegraphics[width = \columnwidth]{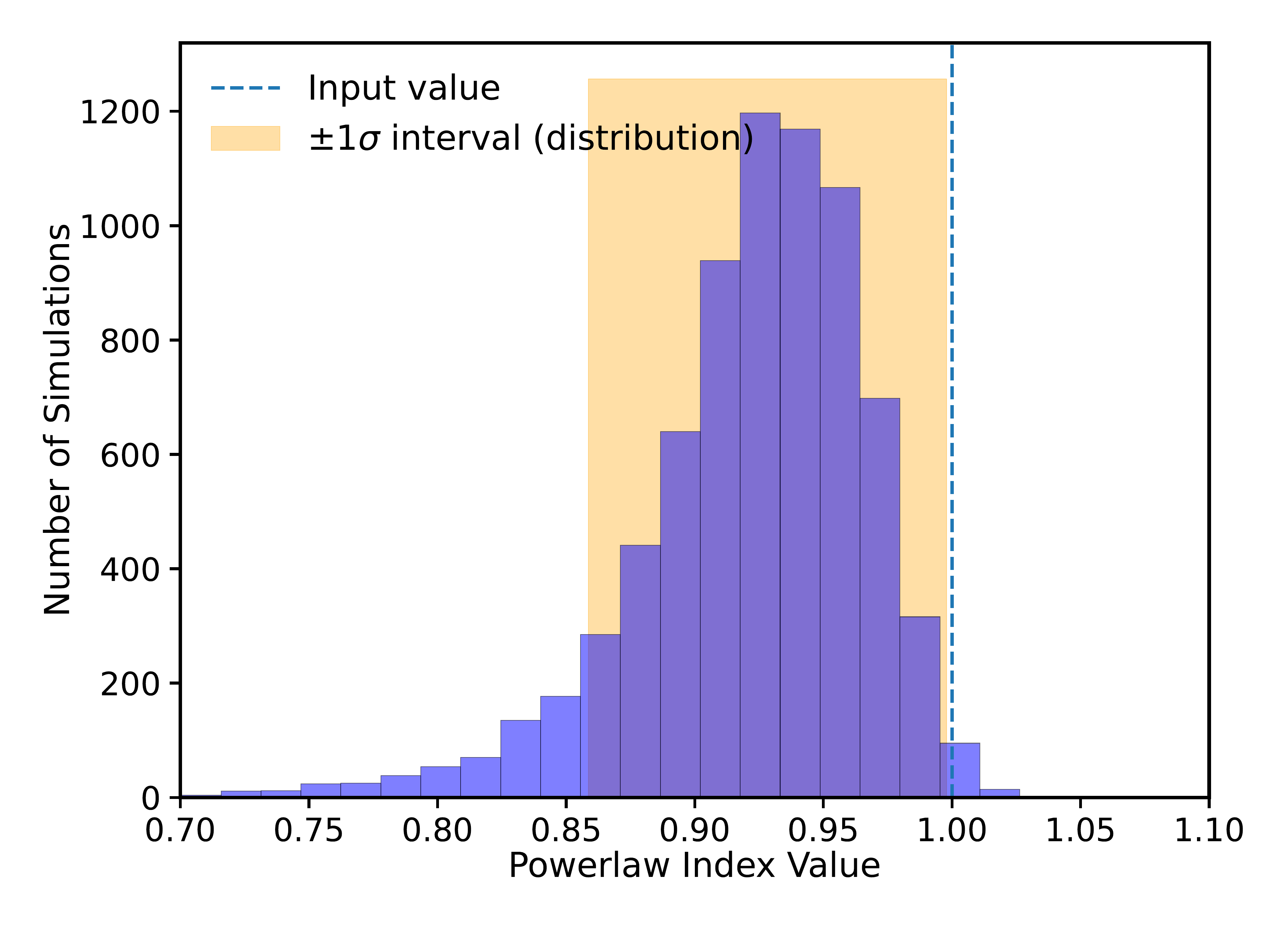}
    \caption{{\bf Check for powerlaw index bias due to uneven sampling.} The distribution represents the powerlaw index values of the simulated LSPs with an input index of 1.0.   \label{fig:lspbias}}
\end{figure}
\begin{figure}[htbp!]
    \centering
    \includegraphics[width = 0.8\columnwidth]{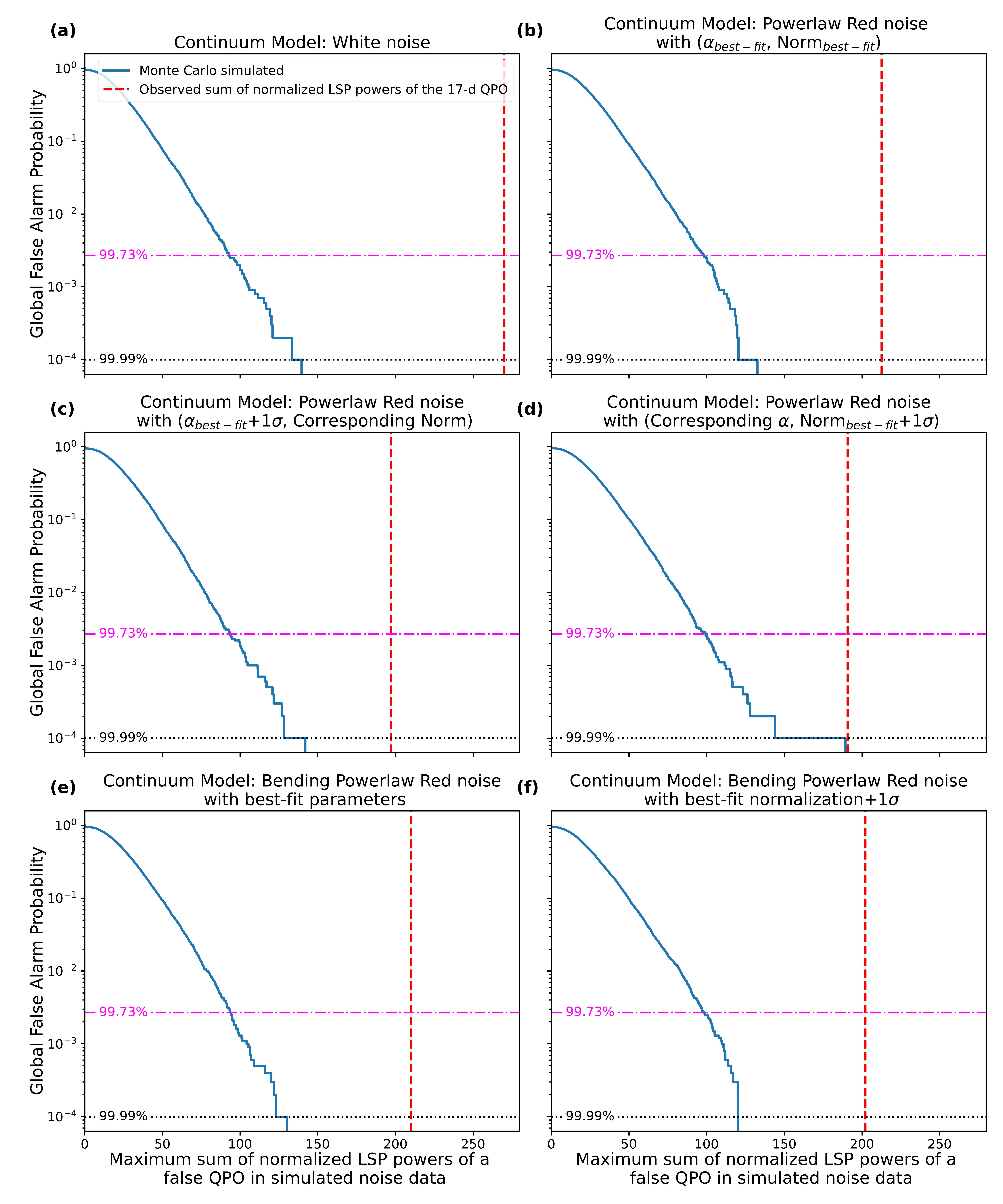}
    \caption{{\bf Estimates of the global false alarm probability (FAP) of finding a broad peak in the simulated LSPs.} Each panel is for a different underlying noise continuum: (a) white noise, (b) powerlaw red noise with best-fit powerlaw index, $\alpha_{best-fit}$, and normalization, Norm$_{best-fit}$, (c) powerlaw red noise with best-fit powerlaw index + 1$\sigma$ and its corresponding normalization, (d) powerlaw red noise best-fit normalization + 1$\sigma$ and its corresponding powerlaw index, (e) bending powerlaw red noise with best-fit parameters, and (f) bending powerlaw red noise with best-fit normalization + 1$\sigma$ and corresponding parameters. The dashed, vertical line in each panel represent the observed QPO at 15-days. 
    \label{fig:fap}}
\end{figure}


\newpage
\begin{figure}[htp!]
\begin{center}
\includegraphics[width=\textwidth, angle=0]{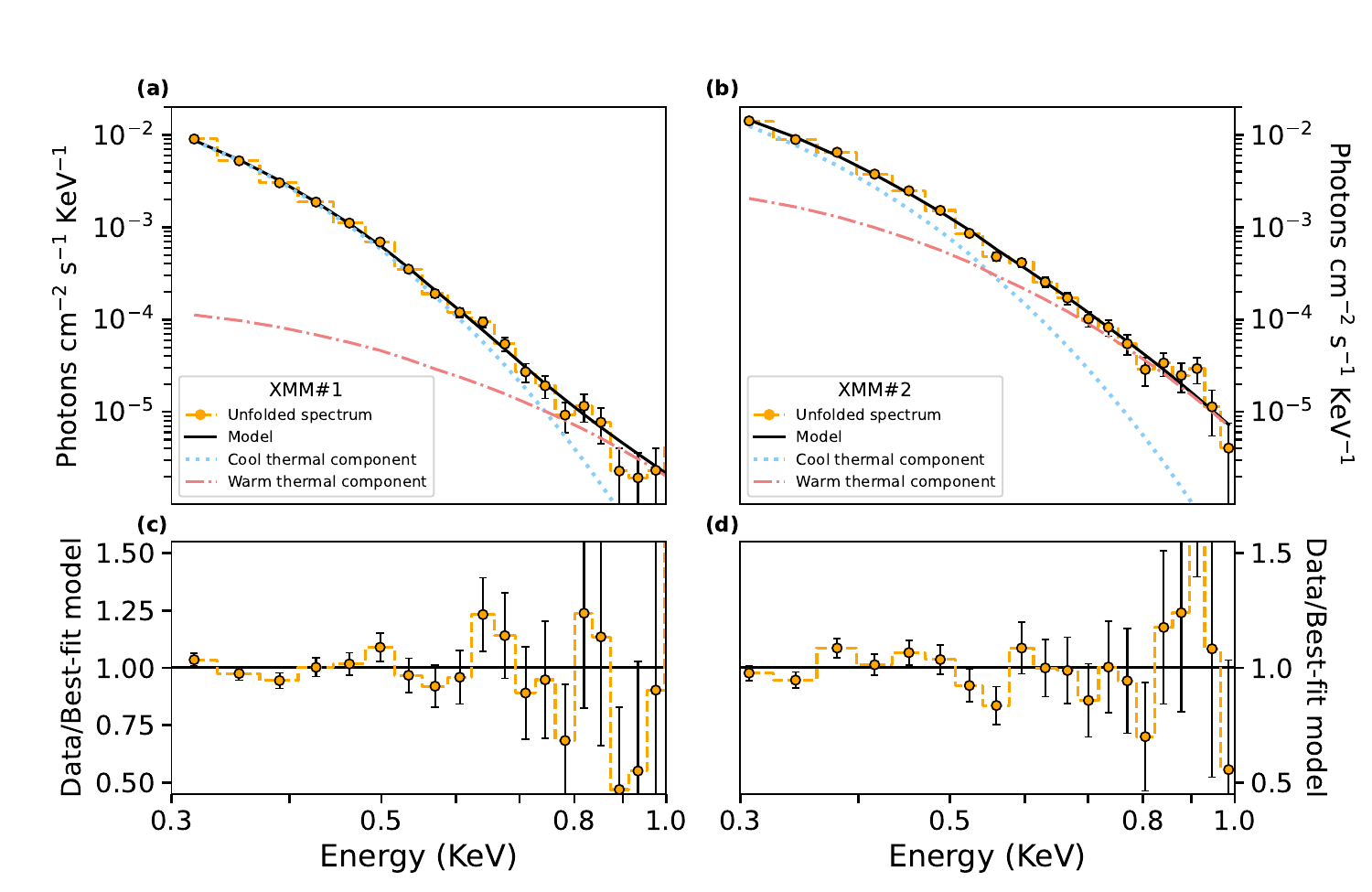}
\end{center}
\caption{{\bf \target's \xmm X-ray spectra.} The best-fit model consists of two thermal components. These spectra are available as supplementary files. All the errorbars represent 90\% uncertainties.}
\label{fig:fig2}
\end{figure}
\vfill\eject


\newpage
\begin{figure}[htbp!]
\begin{center}
\hspace{-1.35cm}
\includegraphics[width=\textwidth, angle=0]{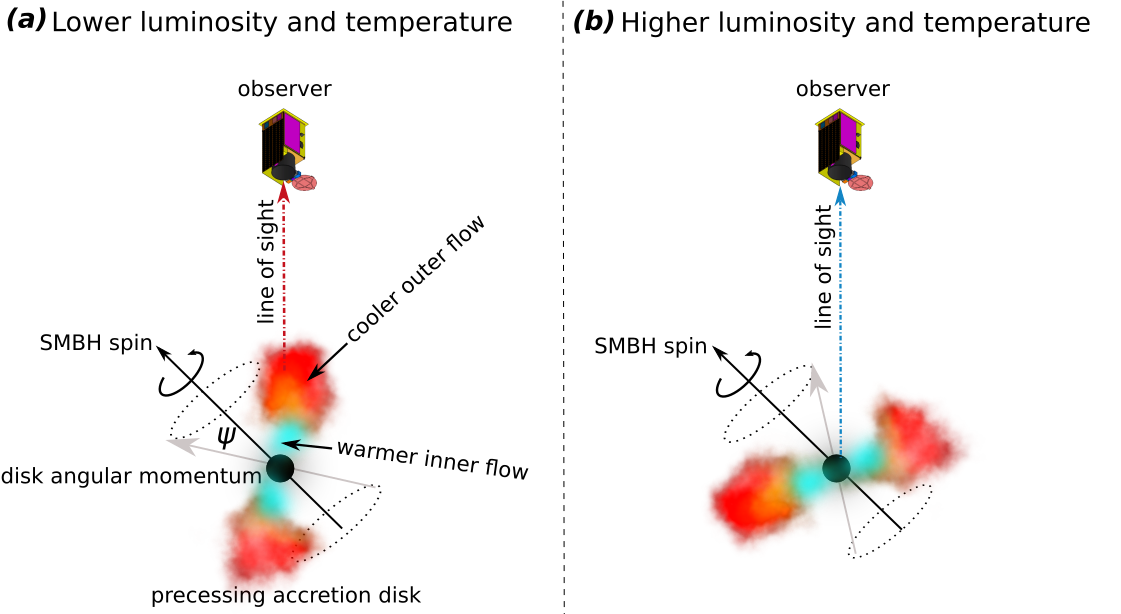}
\end{center}
\vspace{-.35cm} 
\caption{{\bf A simplified schematic of a potential model showing Lense-Thirring precession of an inner disk.} In the left precession phase a), view of the inner/warm disk is obstructed and, consequently, we would see lower luminosity and temperature. In the right panel b), the inner/warm disk is visible which leads to higher observed temperature and luminosity. Relative sizes are not to scale.}
\label{fig:scheme}
\end{figure}

\clearpage
\begin{table}[htp!]
    \centering
    \caption{{\bf Summary of spin constraints for \target SMBH for different surface-density slopes of the precessing accretion flow and the range of SMBH masses.} The negative spin values in parentheses correspond to the retrograde spin. The outer radius of the precessing disk is fixed at the corresponding tidal radius for a Sun-like star. For the radial surface-density profile, we use the prescription $\Sigma \propto R^{-s}$. The last line summarizes the spin interval for a given SMBH mass.}
    \begin{tabular}{c|c|c|c}
    \hline
    \hline
     slope   & $M=10^7\,M_{\odot}$ & $M=10^{6.4}\,M_{\odot}$ & $M=10^{5.8}\,M_{\odot}$  \\
     \hline
     $s=-3/2$    & 0.06 (-0.07)   &  0.15 (-0.16)   &  0.50 (-0.53)   \\
     $s=0$       & 0.06 (-0.06)   &  0.08 (-0.09)   &  0.16 (-0.18)  \\
     $s=+3/5$    & 0.05 (-0.06)   &  0.06 (-0.07)   &  0.08 (-0.09)  \\
     $s=+3/4$    & 0.05 (-0.06)   &  0.06 (-0.06)   &  0.07 (-0.08)  \\
     \hline
     $|a|$ & (0.05, 0.07) & (0.06, 0.16) & (0.07, 0.53) \\
     \hline
    \end{tabular}    
    \label{tab:tab1}
\end{table}

\clearpage
\clearpage

\setcounter{page}{1}
\setcounter{section}{0}
\renewcommand{\theequation}{S\arabic{equation}}
\section*{{\Huge Supplementary Materials.}}
\begin{sloppypar}
\section{A note of lack of similar signals in literature}
There are three factors that dictate the possibility of detecting disk precession in TDEs: (1) the time for stellar debris to circularize and form a disk after disruption, (2) the time for the disk to become thin and align with the black hole spin (see, e.g., Fig.~2 of \cite{2012PhRvL.108f1302S}), and finally, 3) the geometric orientation of the system with respect to our line of sight so that the flux density variations are maximized. The circularization timescale marks the beginning of disk precession while the disk alignment time marks its termination. Even if high-cadence observations are made between the above two epochs, a system that is close to face-on will not show any detectable X-ray modulations. Thus, the search for Lense-Thirring precession in TDEs is best-suited for close to edge-on systems which promptly form an accretion disk, i.e., systems that show early X-ray emission. Only a handful of TDEs have been monitored with high-cadence in the X-rays, and this in combination with the geometric constraints might explain the lack of such signatures in previously-known TDEs. 

\section{Discussion of Various Models}\label{sec:models}
\subsection{Partial tidal disruption event}
The lack of emission prior to the initial detection implies that, if the X-ray modulations are generated by a repeating partial TDE, the star must have been placed on a 15-day orbit through some mechanism. Tidal interactions alone cannot dissipate enough energy to yield the required orbit \cite{cufari2022}, and for the black hole mass inferred for \target Hills capture would require a $\sim 1000$-second orbit of the original binary and a sub-solar-radius separation \cite{cufari2022} for solar-like (i.e., one-solar mass) binary components. In addition, the fact that the optical and UV lightcurves show evolution on $\sim 30-50$ day timescales implies that the fallback time is on this order, which would be longer than the 15-day orbit of the star (which is presumably responsible for modulating the X-ray emission). It seems unreasonable for the fallback time to be longer than the orbital time of the star, rendering this scenario extremely unlikely.

\subsection{Variability from discrete stream collisions}
We can expect that some variability in the TDE accretion process comes from the dynamics of the returning debris stream. For example, the recurrent X-ray flares could, in principle, result from the infall of material onto the black hole from self-intersection shocks. In particular, recent simulations \cite{sadowski16, andalman22} have shown that the self-intersection of the debris stream from a deep TDE leads to a geometrically inflated, slowly evolving, quasi-spherical flow at large radii that can extend to 100's-1000's of gravitational radii. However, on small scales and very near the horizon of the black hole, where the $\sim$ keV emission would originate, \cite{andalman22} found that the flow and the accretion is modulated on timescales comparable to the freefall time from the self-intersection shock, as material dissipates energy and periodically falls to the black hole on this timescale. Geometrically it follows that the self-intersection radius $r_{\rm SI}$ -- assuming that the self-intersection arises from the general relativistic advance of periapsis -- is related to the pericenter distance of the star $r_{\rm p}$ via
\begin{equation}
  r_{\rm SI} = \frac{r_{\rm p}^2}{3\pi GM/c^2}, \label{rSI}
\end{equation}
where $M$ is the mass of the black hole. If we associate $15$ d with the freefall time $t_{\rm ff}$ from the self-intersection shock, then $r_{\rm SI} \simeq \left(\sqrt{GM}t_{\rm ff}\right)^{2/3}$, and letting $M = 10^6M_{\odot}$, Equation \eqref{rSI} yields a pericenter distance of $r_{\rm p} \simeq 200 GM/c^2$. This distance is a factor of a few larger than the classical tidal disruption radius of a solar-like star by a $10^6M_{\odot}$ black hole, and the self-intersection radius is $r_{\rm SI} \simeq 4400 GM/c^2$, i.e., highly spatially extended compared to a geometrically thin disk at the tidal radius.

In this scenario, the material from the self-intersection shock would fall toward the black hole to form a small-scale disk. This disk could generate the higher temperature X-ray emission (i.e., the warmer component revealed by the X-ray analysis of \target). The higher disk temperature, and thus additional pressure support, could inflate the disk to the point that it once again obscures our line of sight, resulting in the rapid shutoff of the X-ray emission. However, to make the above estimate of the pericentre distance that is required to generate the $17$ d timescale commensurate with the tidal radius of a solar-like star requires the black hole mass to be $\sim 10^5\,M_\odot$, which is smaller than that implied by the $M-\sigma_{*}$ scaling for \target.

Another possibility is the variability induced from the stream colliding with the accretion disk. In the case that the disk is misaligned to the black hole spin and precessing, the radius at which the stream hits the disk varies with time. This, in turn, affects the accretion rate through the disk as the angle between the stream and disk angular momentum varies. This affect arises both through the radius at which mass is added to the disk, and through the amount of angular momentum cancellation that is caused by the addition of material with a roughly constant angular momentum direction to a disk with a time-varying angular momentum direction. As noted in the main text, these complications could plausibly explain the more erratic behavior of the X-ray lightcurve than can be explained by a precessing, planar disk, but we leave a detailed investigation of these features to future work.

\subsection{Radiation-pressure instability}\label{supsec:rpi}
Inner portions of standard thin disks are unstable when dominated by radiation pressure \cite{1974ApJ...187L...1L}. This can eventually lead to quasi-periodic flares in the accretion rate explaining the behavior of some changing-look AGN. The model proposed by \cite{2020A&A...641A.167S} consists of an inner advection-dominated hot flow, which is a stable optically-thin solution, and the outer standard thin disk, whose inner zone with the width of $\Delta R$ is radiation-pressure dominated and located on an unstable branch in the accretion rate-surface density plane. The flares due to radiation-pressure instability recur on the timescale, $\tau_{\rm flare}$, shorter than the standard viscous timescale at the distance $R$, i.e. $\tau_{\rm flare}=\tau_{\rm visc} \Delta R/R$, where $\tau_{\rm visc}$ is the viscous timescale that depends on the distance $R$, the viscosity $\alpha$ parameter, and the flow scale-height $H$. The unstable zone forms for a relative (Eddington) accretion rate larger than $\dot{m}\gtrsim 0.15(\alpha/0.1)^{41/29}(M/10^6 M_{\odot})^{-1/29}$ \cite{2020A&A...641A.167S}, which can be achieved in \target for $M\lesssim 10^{6.7}\,M_{\odot}$. 

As was shown by \cite{2023A&A...672A..19S}, via global time-dependent calculations without the assumption of the hot inner {\bf advection-dominated accretion flow (ADAF)}, the model can be fine-tuned enough, including corona and sufficiently strong magnetic field, to produce quasiperiodic flux changes by an order of magnitude on the timescale as short as $\sim 10$ days, when the disk is small enough (like for TDEs) or interrupted by a gap. For the early phases of the TDE, we expect Eddington to super-Eddington accretion rates and in that case, the accretion disk is expected to be a geometrically thicker slim disk that is stabilized by 
advection. The radiation-pressure instability model is still applicable for a standard disk, and the advective terms are included in the calculations via radial derivatives of density and temperature
(time-dependent numerical simulations tailored to \target will be presented in a separate study). For \target this is relevant  if its SMBH is heavier, e.g. for $M\sim 10^7\,M_{\odot}$, i.e., if the Eddington ratio is $\dot{m}\sim 0.1$, so that the accretion disk can be of a standard type. The model is applicable as well for higher accretion rates. However, a challenge to the instability model is the fact that \target's flares weaken after a few months (Fig. \ref{fig:fig1}a). This can happen if the magnetic field gets sufficiently strong, e.g. due to  dynamo effect,
which tends to stabilize the disk \cite{2023MNRAS.524.1269K}. In that case we would expect the amplitude to gradually decrease with the magnetic field build-up. 
 
\subsection{Disk tearing}\label{supsec:disktearing}
In the main text we discussed the response of the TDE disk to Lense-Thirring precession in the case that the disk is sufficiently hot so that warp waves, which propagate at a velocity of roughly half the sound speed, are able to communicate the precession efficiently to create rigid precession of the (warped) disk. If the disk is thin enough such that it develops a strong warp then it can be unstable to disk tearing, in which the disk breaks apart into discrete rings that precess at roughly the local Lense-Thirring rate \cite{Nixon2012,raj2021}. The disk tearing instability is understood from an analytical standpoint \cite{Ogilvie2000,Dogan2018,Dogan2020}, and the associated nonlinear behaviour has now been explored in a variety of numerical simulations (including Lagrangian and Eulerian codes; e.g. \cite{Nixon2012,Liska2021} and in both the high- and low-viscosity regimes \cite{Drewes2021}. Recently, a protoplanetary disk in the multi-stellar system GW Ori has been seen to harbour misaligned and broken rings of gas using spatially resolved observations, and this has been ascribed to the disk tearing process \cite{Kraus2020} (where, in this case, the precession was driven by the gravitational torque from the orbiting stars; \cite{Nixon2013}). The variability induced by the disk tearing process can be due to both geometric effects (e.g. rings precessing through the line of sight) or intrinsic disk variability (e.g. enhancement of the central accretion rate due to rapid accretion from interacting rings). The timescale for the variability is of the order of the local Lense-Thirring precession timescale in the unstable region of the disk \cite{Raj2021b}. For the timescale to be $\approx 15.9$\,days this would suggest rings precessing at radii of $R_{\rm prec} \approx 35\,R_{\rm g}\,(a/0.5)^{1/3}M_6^{-1/3}$, where $M_6 = M/10^6\,M_\odot$. Using the simple estimate provided by \cite{Nixon2012} for where we would expect the disk to break into discrete rings, we have $R_{\rm break} \approx 15 R_{\rm g}\,(a/0.5)^{2/3}(\alpha/0.1)^{-2/3}(H/R\,/\,0.1)^{-2/3}$, where $\alpha\approx 0.1$ \cite{Martin2019} is the Shakura-Sunyaev disk viscosity parameter and $H/R$ is the disk angular semi-thickness. It is therefore possible to find reasonable parameters that are consistent with what we know about \target to bring these into agreement. For example, with a black hole mass of $10^{6.4}M_\odot$, a spin of $a \approx 0.5$, $\alpha = 0.1$, and $H/R \approx 0.05$ (consistent with, e.g., the disk model of \cite{Strubbe2009} with ${\dot M}/{\dot M}_{\rm Edd} = 0.3$), we can expect disk tearing to occur at a radius of $\approx 25 R_{\rm g}$ with a period of $\approx 15$\,days. Thus it is possible that the variability observed in the X-ray lightcurve from \target is due to either Lense-Thirring precession of a rigidly precessing and radially extended portion of the disk (as discussed in the main text) or Lense-Thirring precession of an unstable region of the disk that has broken into discrete rings.

\end{sloppypar}


\putbib
\end{bibunit}
\end{document}